# Wave-ice interaction in the North-West Barents Sea

A. Marchenko[1], P.Wadhams[2], C.Collins[3], J.Rabault[4], M.Chumakov[5]

[1] The University Centre in Svalbard, Longyearbyen, Norway
[2] Univetrsity of Cambridge, Cambridge, UK
[3] U.S. Army Engineer Research and Development Center, Coastal and Hydraulics Laboratory, Duck, NC, USA
[4] University of Oslo, Oslo, Norway
[5] Gazprom VNIIGAZ LLC, Moscow, Russia

**Abstract**

The results of field work on drift ice during wave propagation are analyzed and presented. The field work was performed in the Barents Sea, and the main focus of the paper is on wave processes in the MIZ. A model of wave damping in broken ice is formulated and applied to interpret the field work results. It is confirmed that waves of higher frequencies are subjected to stronger damping when they propagate below the ice. This reduces the frequency of most energetic wave with increasing distance from the ice edge. Difference of wave spectra measured in two relatively close locations within the MIZ is discussed. The complicated geometry and dynamics of the MIZ in the North-West Barents Sea allow waves from the Atlantic Ocean and south regions of the Barents Sea to penetrate into different locations of the MIZ.

1. ## Introduction

Penetration of surface waves below floating ice and their action on the ice are recently of interest because of a reduction in ice covered areas in the Arctic Ocean. Waves can destroy relatively big areas of solid drift ice in few hours (see, e.g., Collins et al, 2015). Broken ice is more mobile and has less insulating capacity for energy exchange between atmosphere and ocean. Wave amplitudes and lengths are the most important characteristics influencing critical bending stresses in ice leading to its eventual destruction. The process of wave attenuation controls wave amplitudes and spectral shape, and it depends on the distance to the ice edge.

Wadhams et al. (1988) investigated characteristics of wave attenuation process in the marginal ice zone (MIZ) of the Greenland Sea and Bering Sea. Similar processes have been investigated in the Barents Sea by Frankenstein et al (2001), Tsarau et al (2017), Marchenko and Chumakov (2017), and in the Antarctic ice by Meylan et al (2014) and Doble et al (2015). Field measurements have demonstrated exponential attenuation of wave amplitudes along their propagation direction into ice covered areas. Attenuation of wave amplitude and wave energy increases with the increasing wave frequency. The dependence of wave attenuation from wave amplitudes is not significant in most of the cases.

Physically, wave attenuation in ice covered ocean areas is associated with the scattering of wave energy by floe edges, anelastic deformations of floes, floe-floe interactions, energy dissipation in water boundary layer under the ice, floe-slush interactions, and wave induced migration of brine through the ice. Scattering of wave energy on floe edges has been investigated in numerous papers (see, e.g., Squire, 2007; Meylan and Masson, 2006). Viscoelastic model of ice is considered by Wang and Shen (2010). Wave damping due to energy dissipation in under ice boundary layer was formulated by Weber (1987). Wave damping due to floe-floe interaction in broken ice was considered by Keller (1998). A comparison of wave attenuations caused by anelastic bending deformations of solid ice, energy dissipation in under ice boundary layer and wave induced migration of brine is discussed in Marchenko and Cole (2017).

In the paper, the results of field work on drift ice during wave propagation are analyzed and presented. The field work was performed in the Barents Sea, and the main focus of the paper is on wave processes in the MIZ. In the second section, we describe the field work. Sections 3 and 4 are devoted to the description and analysis of the results. Section 5 describes an analysis of high resolution SAR-image of the region where the field works was carried out. A model of wave damping in broken ice is formulated in Section 6. Section 7 is devoted to a discussion of the results.

2. **Locations and organizing of field works**

Field work was performed on the drifting ice near Edgeøya in the Nort-West Barents Sea on May 01 and May 05 of 2016. Ice conditions during this period are shown on ice maps in Fig. 1. From April 29 to May 02 (until 03:00 UTC) the South-West and West wind influenced ice drift in the North-East and East directions. The wind caused ice divergence near Edgeøya and displacement of the ice edge to the East. Then the wind direction changed to North-East and it influenced convergence of ice near Edgeøya. The wind direction and speed are shown in Table 1 constructed using the data by https://earth.nullschool.net.

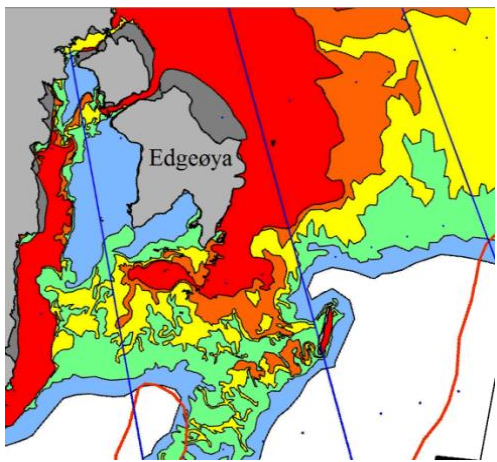  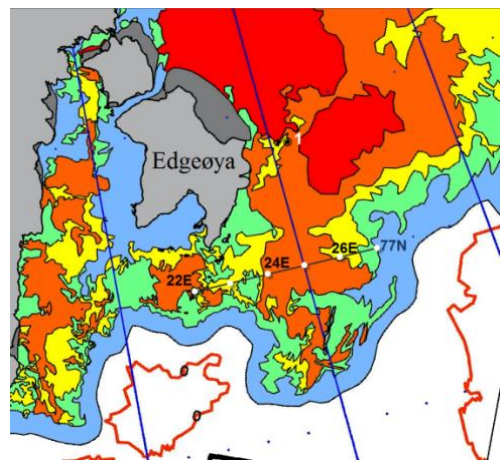

April 29                                  May 02

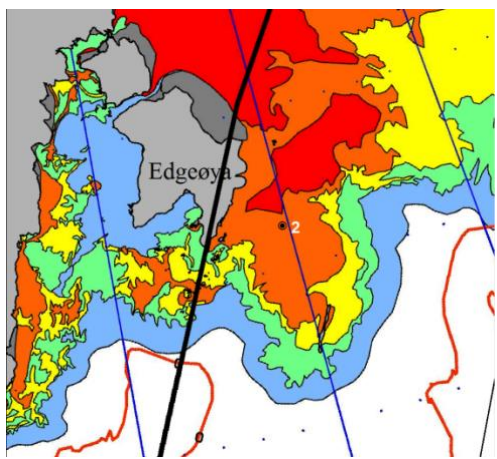 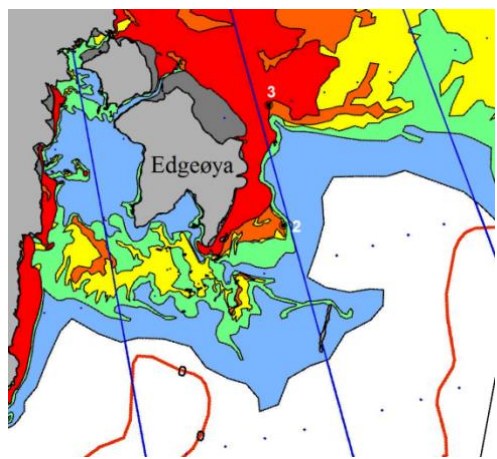

May 04                                              May 06

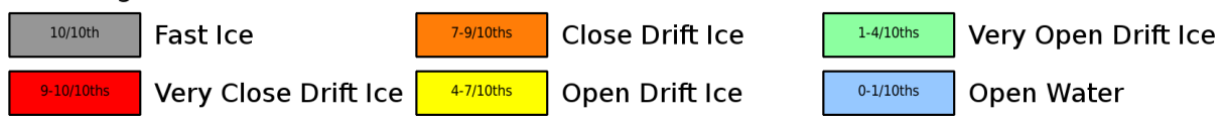

Figure 1. Ice conditions from April 29 to May 06 of 2016. Points 1 and 2 mark the locations of field works on May 01 and May 05. Point 3 show the location of the SAR image.

Table 1. Wind velocity at 12:00 UTC from April 29 to May 06 of 2016.

|  | Apr. 29 | Apr. 30 | May 01 | May 02 | May 03 | May 04 | May 05 | May 06 |
|---|---|---|---|---|---|---|---|---|
| Speed, km/h | 11 | 18 | 16 | 6 | 18 | 19 | 31 | 30 |
| Direction, from | 230 | 215 | 235 | 60 | 135 | 130 | 110 | 105 |

On the station 1, the RV Lance was moored to a floe with a thickness around 30 cm and a diameter around 2 km on May 02 (Fig. 2a). Geographical location was 25.5ºE, 77.76ºN. Sea depth in the region was measured to be around 160 m. The floe mass was estimated between $7 \cdot 10^5$ tons and $8 \cdot 10^5$ tons. The mass of fully loaded Lance is $M_L$=2370 tons, and her length and breadth are around $L_L$=60 m and $w_L$=12 m. Therefore, the influence of the Lance on the floe dynamics would have been negligible. Frequency of natural oscillations of the Lance is estimated by the formula $\omega_L^2 = \rho_w g S_L / M_L$, where $\rho_w$ is the water density, $g$ is the gravity acceleration, and $S_L = w_L L_L$. The natural frequency is of $\omega_L \approx 1.7$ rad/s, and the period is of $T_L \approx 3.5$ sec. This period is smaller than typical swell periods of 8-12 s registered in the Barents Sea (see, e.g., Marchenko et al, 2015).

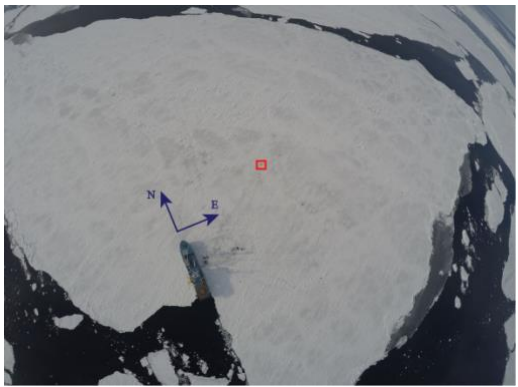
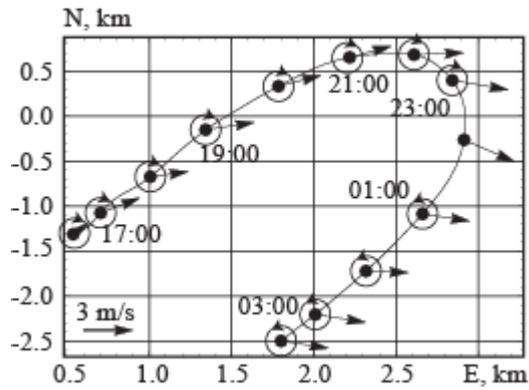

a)                                       b)

Figure 2. (a) View of RV Lance from drone. Red square shows location of wave measurements. (b) Trajectory of ice tracker from 17:00, May 01, to 03:00, May 02. Vectors of the wind velocity measured along the drift trajectory are shown by rectilinear arrows. Circular arrows show the direction of floe rotation.

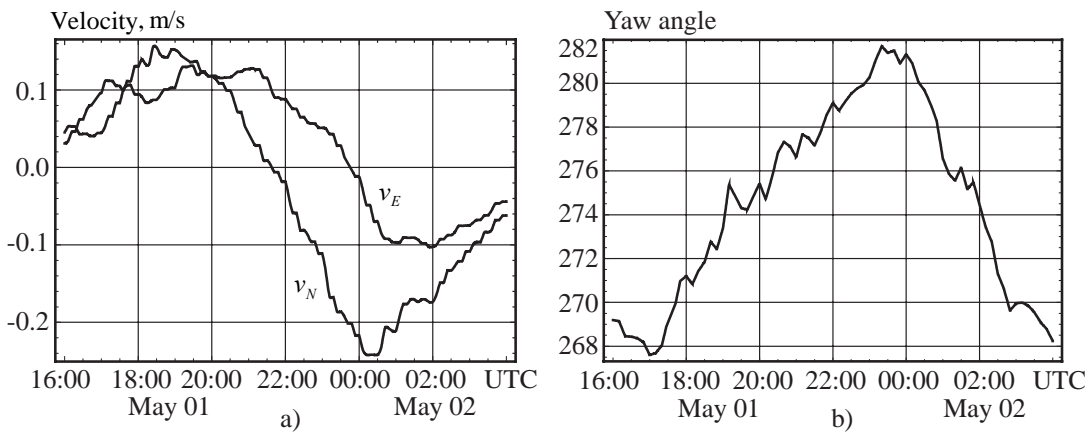

Figure 3. The East and North components of the floe velocity (a) and yaw angle (b) versus the time reconstructed from the ice tracker records.

Equipment was deployed on the floe in the afternoon of May 01, 2016, and wave measurements started from 16:00 (here and further UTC is used), May 01, and extended until 09:00, May 02. The floe drift and wind velocity were registered with an Oceanetic Measurements Ice Tracker (model 703 equipped with anemometer, magnetometer and compass). Measurements of wave characteristics were performed under the ice with the pressure and temperature recorders SBE-39 (Sea-Bird Scientific) and an Acoustic Doppler Velocimeter SonTek Ocean Probe 5 MHz (ADV), and on the ice surface with accelerometers in a custom housing and recording unit fabricated at the University of Oslo (Sutherland et al, 2017).

The Ice Tracker was deployed on the floe around 11:00 and telemetered data via Iridium. The Ice Tracker trajectory during the time of wave measurements is shown in Fig. 2b. Wind velocity vectors are shown by rectilinear arrows in the same figure. Circular arrows show the direction of the floe rotation. The floe rotated in clockwise direction on approximately 10º from 16:00 to 24:00, May 01, and on counter clockwise direction on approximately 15º from 00:00 to 06:00, May 02. The East and North components of the floe velocity and yaw angle reconstructed from the ice tracker records are shown in Fig. 3 versus the time.

Deployment location of SBE-39 and ADV is shown in Fig. 2a by red square. The accelerometers were deployed along the line crossing the red square and extended in the North direction. Two recorders SBE 39 were mounted on a steel wire on the depths of 3.6 m and 11.4 m, and the wire was fixed on the ice surface. The ADV probe is equipped with tilt sensor and compass. Magnetic inclination about 15º in the region of the field works was taken into account in the data processing. The ADV was mounted on the vertical wooden pole fixed on the ice by a tripod. Water velocity measurements were performed on the depth 80 cm below the ice. The depth was measured by the ADV pressure sensor. The SBE-39 recorders were deployed around 16:00, May 01, and recovered around 09:00, May 02. The ADV sensor was deployed on 19:00, May 01. It was disconnected from the electric power source on the Lance board around 04:30, May 02, because of the floe movement relative to the ship.

Custom accelerometer unit manufactured in the University of Oslo were equipped with VN-100 [23]. The VN-100 is a miniature, high-performance Inertial Measurement Unit (IMU) and Attitude Heading Reference System (AHRS). The VN-100 combines 3-axis accelerometers, 3-axis gyros, 3-axis magnetometers, a barometric pressure sensor and a 32-bit processor. Henceforth, the name IMU is used instead accelerometer. 10 IMUs were deployed on the floe around 17:00, May 01, and recovered around 09:00, May 02. The positions of the 10 IMUs deployed are presented in Fig. 4b. One IMU was deployed alone, while all other IMUs were grouped into arrays of three sensors.

Flexural strength of sea ice on the station 1 was measured in two tests with floating cantilever beams loaded at the end in downward direction, and two tests with floating cantilever beams loaded at the end in upward direction. The tests results are shown in Table 2, where $l$, $b$ and $h$ are the beam length, width and thickness, $T$ and $S$ are the temperature and salinity of the ice averaged over the thickness, $F_{max}$ and $\delta_{max}$ are the maximal loads and maximal displacements of the beams ends, $\sigma_f$ is the flexural strength, and $E$ is the effective elastic modulus calculated from the beam theory. Values of flexural strength and elastic modulus were calculated and are in the range of values specified by the formula of Timco and Brien (1994) and other investigations of flexural strength of sea ice in the Barents Sea and Svalbard region (see, e.g., Marchenko et.al., 2017a).

Table 2. Results of bending tests with cantilever beams.

| N | LD | $l$, cm | $b$, cm | $h$, cm | $T$,ºC | $S$, ppt | $F_{max}$, N | $\delta_{max}$, mm | $\sigma_f$, MPa | $E$, GPa |
|---|------|-----|----|----|----|------|------|------|------|------|
| 1 | Down | 184 | 39 | 45 | -2 | 4.78 | 2317 | -    | 0.32 | -    |
| 2 | Down | 220 | 40 | 29 | -2 | 4.78 | 787  | 2.6  | 0.31 | 1.32 |
| 3 | Up   | 184 | 42 | 27 | -2 | 3.71 | 1004 | -1,6 | 0.36 | 1,89 |
| 4 | Up   | 215 | 37 | 24 | -2 | 3.69 | 515  | -2,3 | 0.31 | 1,74 |

On station 2, Lance was drifting between floes with diameters of 3-5 meters in energetic wave conditions. Sea depth was around 280 m. The rescue boat was used to deploy and recover IMUs on four floes (Fig. 4a). Three IMUs were deployed on neighbor floes and one IMU was deployed on a floe located in 20-30 m from the three floes (Fig. 4b). The records were performed during two hours from 09:00 to 11:00, May 05. The Lance was drifting few hundreds of meters down-wave from the IMUs.

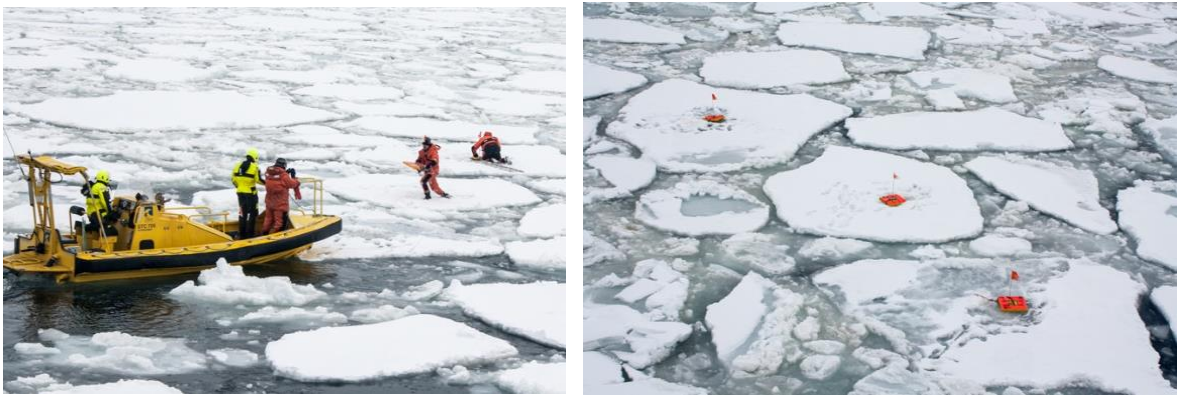

a)                             b)

Figure 4. Deployment of IMUs of the drift ice (a). Three IMUs on neighbor floes (b).

### 3. Results of field measurements on station 1

Analysis of collected data shown similar wave characteristics as measured by SBE-39, ADV and IMUs sensors (Marchenko et al., 2017b). Therefore, we confine additional analysis to the SBE and ADV measurements. User setups for the SBE and ADV measurements are given in Table 3. Figure 5 shows spectrogram and the absolute values of the Fourier transform $\delta p_{3,f}(\omega)$ of the water pressure fluctuations $\delta p_3(t)$. The water pressure fluctuations were extracted from the water pressure recorded by SBE 39 in decibars (depth(meters)=pressure(decibars)*1.02) at 3.6 m depth to avoid the influence of external sources (changes of atmospheric pressure, sliding of the wire, etc) on the fluctuations caused by hydrodynamics effects. They were calculated using Wolfram Mathematica 11.2 software. The spectrum has four local maxima at frequencies 0.06 Hz, 0.08 Hz, 0.1 Hz and 0.125 Hz before 23:00, May 01, and there are only two local maxima at frequencies 0.06 Hz and 0.09 Hz after 00:00, May 02. The frequency of waves with maximal energy was 0.125 Hz before 23:00, May 01, and 0.09 Hz after 00:00, May 02. Two peaks close to each other on blue line in Fig. 5b correspond to wave modulations.

Table 3. User setups of ADV and SBE-39. SPB – number of samples per burst, NB – number of bursts.

| Sensor | Sampling frequency | Burst Interval | SPB | NB | Depth |
|--------|--------------------|----------------|------|------|-------|
| ADV    | 10 Hz              | 360 s          | 2400 | 94   | 0.8 m |
| SBE 39 | 1 Hz               | -              | -    | -    | 3.6 m |

| SBE 39plus | 2 Hz | - | - | - | 11.4 m |

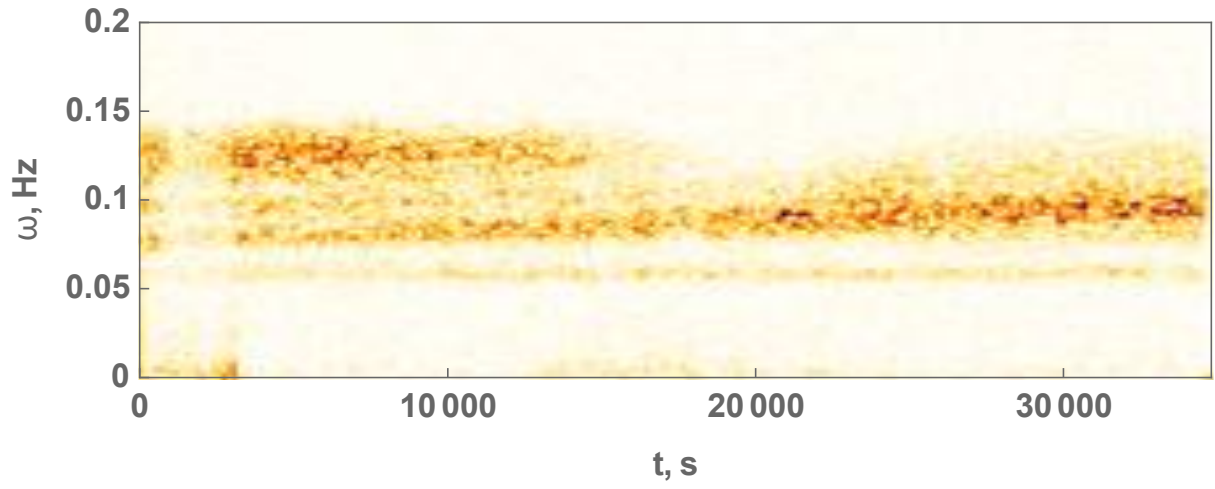

a)

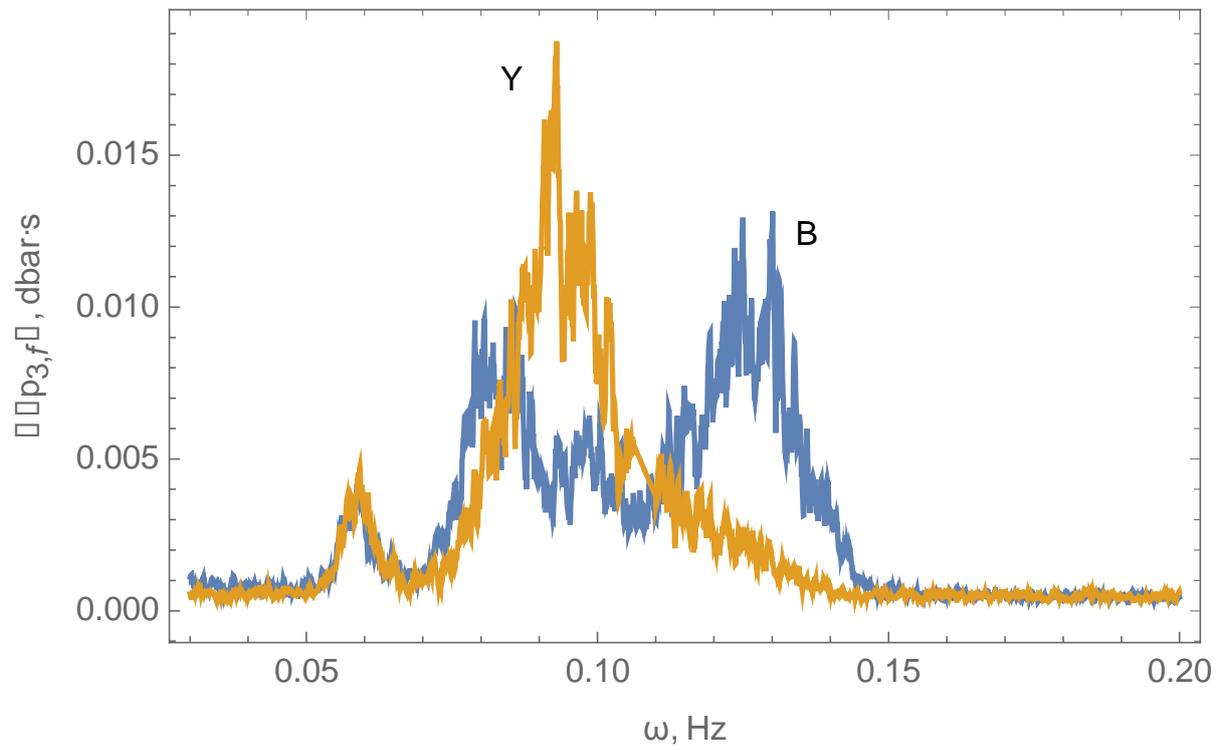

b)

Figure 5. Spectrogram (a) and spectrums (b) of the fluctuations of water pressure recorded on 3.6 m depth. The time is accounted from 16:00, May 01 (a). The spectrums correspond to the records from 19:00 to 23:00, May 01, (blue line B) and from 23:00, May 01, to 03:00, May 02, (yellow line Y) (b).

The ratio of water pressure amplitudes at depths $z=z_1$ and $z=z_2$ caused by periodic gravity wave is equal to (Marchenko et al., 2013)

$$\gamma \equiv \frac{p_{1,dyn}}{p_{2,dyn}} = \frac{\omega^2 \cosh[k(z_1+H)] - gk \sinh[kH]}{\omega^2 \cosh[k(z_2+H)] - gk \sinh[kH]}. \tag{1}$$

Formula (1) is used for the calculation of the wave number $k$ when the wave frequency is known. We consider the Fourier transforms $\delta p_{3,f}(\omega)$ and $\delta p_{11,f}(\omega)$ of the pressure fluctuations $\delta p_3(t)$ and $\delta p_{11}(t)$ recorded at depths 3.6 m and 11.4 m, and assume that the ratio $\gamma$ equals to the ratio $|\delta p_{3,f}|/|\delta p_{11,f}|$, where the values of $|\delta p_{3,f}|$ and $|\delta p_{11,f}|$ are taken in the points of spectral maxima. Thus, the wave numbers corresponding to the spectral maxima are calculated from equation (1). Fig.6a show the values $(k, \omega)$ corresponding to the spectral maxima. Fig.6a shows the dispersion curve of the gravity waves propagating in the water with free surface with the depth of 160 m. It is obvious that the measured values of $(k, \omega)$ sit on the dispersion curve. This means that the influence of ice on dispersion properties of observed waves is negligible for the range of frequencies considered.

This conclusion follows from the dispersion equation of flexural-gravity waves (see, e.g., Greenhill, 1886) $\omega^2 = gk \tanh(kH)(1 + Dk^4)$ if $Dk^4 \ll 1$. Here $D \approx E h_i^3/(12\rho_w g)$, $E$ is the effective elastic modulus of the ice and $h_i$ is the ice thickness. In-situ tests on flexural strength of floating cantilever beams performed on the drift ice on May 01 showed that $E = 1.3 - 1.9$ GPa (Table 2). Assuming $h_i = 0.3$ m we find that $Dk^4 \approx 0.01$ when $k = 0.07$ m$^{-1}$, i.e. the influence of ice elasticity on the waves with registered frequencies is very small in comparison with the gravity force.

Wave amplitude is calculated using water pressure records at depth $z$ below the ice and the formulas:

$$\delta p_{z,f}(\omega) = \rho_w g \eta_f(\omega) P(k,z), \quad P(k,z) = \frac{\cosh[k(z+H)]}{\cosh[kH]} - 1, \tag{2}$$

where $\delta p_{z,f}$ and $\eta_f$ are the Fourier transforms of the water pressure fluctuations $\delta p_z(t)$ at the depth $z$ and water surface elevation $\eta(t)$ caused by waves. We used records of the water pressure at 11.4 m depth. Values of $\delta p_{z,f}$ were calculated using the discrete Fourier transform in Wolfram Mathematica 11.2 Software. Then the inverse Fourier transform was used to calculate water surface elevation as a function of the time separately. The procedure was realized independently for each hour of the record. Significant wave height (SWH) is calculated using the formula

$$\text{SWH} = 4 \cdot \text{StandardDeviation}[\eta]. \tag{3}$$

The dependence of SWH from the time is shown in Fig. 6b. The time is accounted from 00:00, May 01. Mean value of SWH was found around 10 cm. Maximal bending stress in the ice deformed by periodic sinusoidal wave is estimated with the formula $\sigma_{max} = 0.5Eh_i ak^2$, where $a$ is the wave amplitude. Assuming $a = 0.1$ m we find that $\sigma_{max} < 0.15$ MPa when $k < 0.07$ m$^{-1}$ and $E < 1.9$ GPa. Thus, maximal ice stresses excited by the observed waves are smaller than half the flexural strength and, therefore, the waves didn't break up the ice.

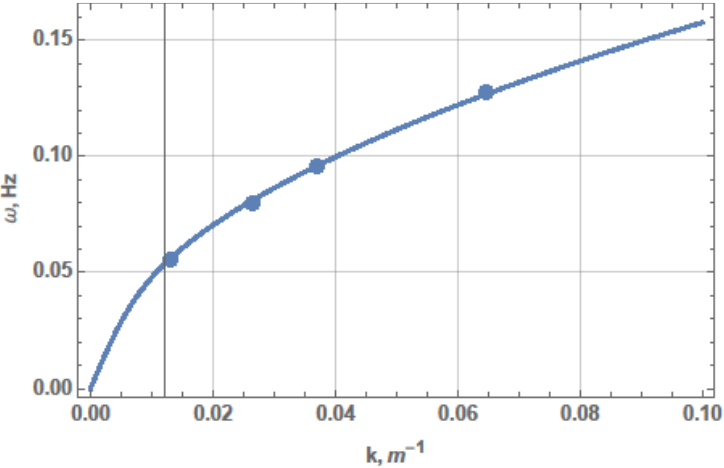

a)

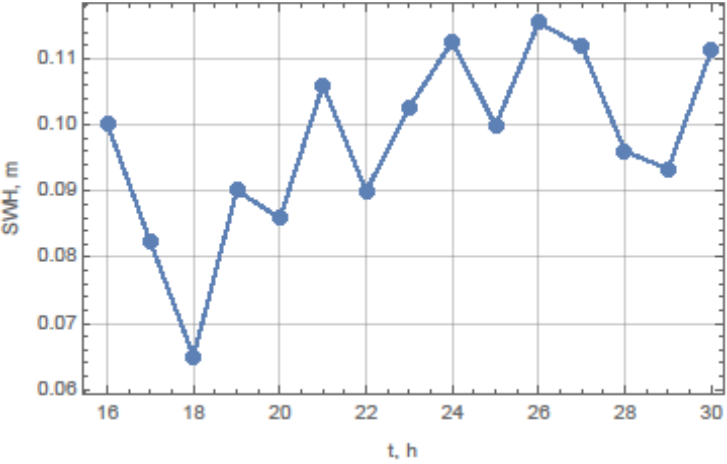

b)

Figure 6. (a) Dispersion curve of surface gravity waves in the water of 160 m depth. Black dots are constructed using SBE-39 records at 3.6 m and 11.4 m depths. (b) Significant wave height versus the time. The time is accounted from 00:00, May 01.

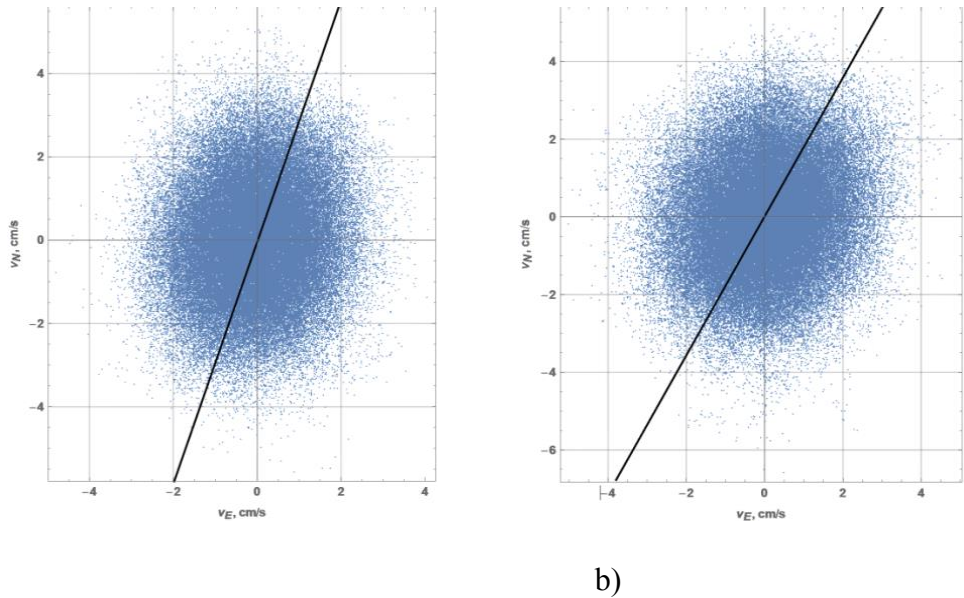

a)  b)

Figure 7. Hodographs of the velocity fluctuations recorded from 19:00 to 23:00, May 01, (a) and from 23:00, May 01, to 03:00, May 02, (b).

Points in Figure 7 show the fluctuations of the horizontal velocities measured by ADV below the ice. They fill regions of elliptic shape. The black lines mark the direction of major axes of the elliptic regions, and point out the direction of propagation of most energetic waves causing highest velocity fluctuations. The black lines are inclined to the North direction under the angles of 18° (Fig. 7a) and 26° (Fig. 7b). The amplitude of periodic wave causing horizontal velocity of surface water particle $v_h$ is estimated in deep water with the formula $\omega a = v_h$, where $a$ is the wave amplitude and $\omega$ is the wave frequency. The frequency of most energetic waves was 0.125 Hz before 23:00, and 0.09 Hz after 23:00 of May 01. The major axes of elliptical regions shown in Fig. 7a and Fig. 7b are estimated respectively as 3.16 cm/s and 3.60 cm/s. The estimated wave amplitude was $a \approx 4$ cm before 23:00, and $a \approx 6$ cm after 23:00 of May 01.

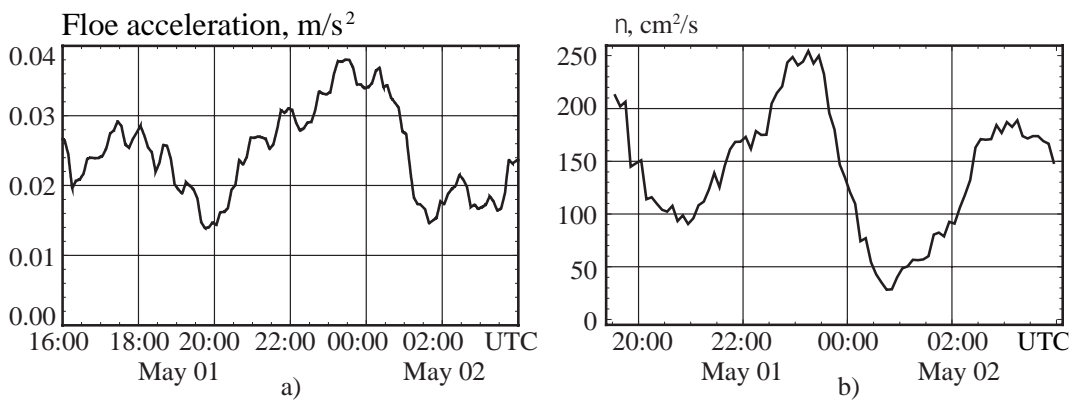

Figure 8. Floe acceleration (a) and the eddy viscosity (b) versus the time.

Figure 8 shows the floe acceleration and the eddy viscosity below the ice versus the time. Correlation between the floe acceleration and the eddy viscosity is visible after 20:00, May 01. Increase or decrease of the acceleration is accompanied respectively by increase or decrease of the eddy viscosity.

The floe acceleration is calculated using the data provided by the ice trackers deployed on the floe. The eddy viscosity is calculated using the data collected in each burst by the formula

$$\nu = \frac{\tau_h h_{ADV}}{\rho_w \langle v_{ADV,h} \rangle}, \tag{4}$$

where $\tau_h$ is the horizontal Reynolds stress, $h_{ADV} = 80$ cm is the depth of the velocity measurements with ADV, $\langle v_{ADV,h} \rangle$ is the mean horizontal velocity measured by ADV, and $\rho_w = 1020$ kg/m³ is the density of sea water. The horizontal Reynolds stress is calculated by the formulas

$$\tau_h = \sqrt{\tau_E^2 + \tau_N^2}, \tau_E = \rho_w \langle v'_E v'_z \rangle, \tau_N = \rho_w \langle v'_N v'_z \rangle, \tag{5}$$

where $v'_E$, $v'_N$ and $v'_z$ are the fluctuations of the East, North and vertical components of the water velocity measured by ADV.

## 4. Results of field measurements on station 2

The acceleration rates are calculated with the formulas (Sutherland et al., 2017)

$$\dot{a}_X = \dot{a}_{m,X} - g\omega_{m,Y}, \dot{a}_Y = \dot{a}_{m,Y} + g\omega_{m,X}, \dot{a}_Z = \dot{a}_{m,Z}, \tag{6}$$

where $\dot{a}_{m,X}$, $\dot{a}_{m,Y}$, and $\dot{a}_{m,Z}$ are the acceleration rates of accelerations $a_{m,X}$, $a_{m,Y}$, and $a_{m,Z}$ measured by IMU along the IMU's axes $X$ and $Y$, $\omega_{m,X}$ and $\omega_{m,Y}$ are the angular velocities relative to the axes $X$ and $Y$ measured by IMU. The acceleration rates $\dot{a}_{m,X}$, $\dot{a}_{m,Y}$, and $\dot{a}_{m,Z}$ are calculated numerically using linear interpolation of measured accelerations over time. The mean values of the accelerations are removed from the data to analyze only wave induced fluctuations. The rate of the horizontal acceleration is calculated with the formula $\dot{a}_h = \sqrt{\dot{a}_X^2 + \dot{a}_Y^2}$, and the rate of the vertical acceleration $\dot{a}_v$ is approximated by $\dot{a}_{m,Z}$.

Figure 9a shows the points at which the coordinates are equal to the rates of the horizontal and vertical accelerations. The points fill a quarter of a circle with a radius of about 0.3 m/s³ and slightly expand outside of the circle. Four spectrums of accelerations normal to the floes are shown in Fig. 9b. All have a peak frequency of about 0.125 Hz and are almost identical. The amplitude of periodic wave causing vertical acceleration rate $\dot{a}_z$ is estimated with the formula $a\omega^3 \approx \dot{a}_z$, where $\omega$ is the wave frequency. Assuming $\omega = 0.78$ rad/s we find $a \approx 0.63$ m. Two peaks closed to each other shown in Fig. 9b correspond to wave modulations.

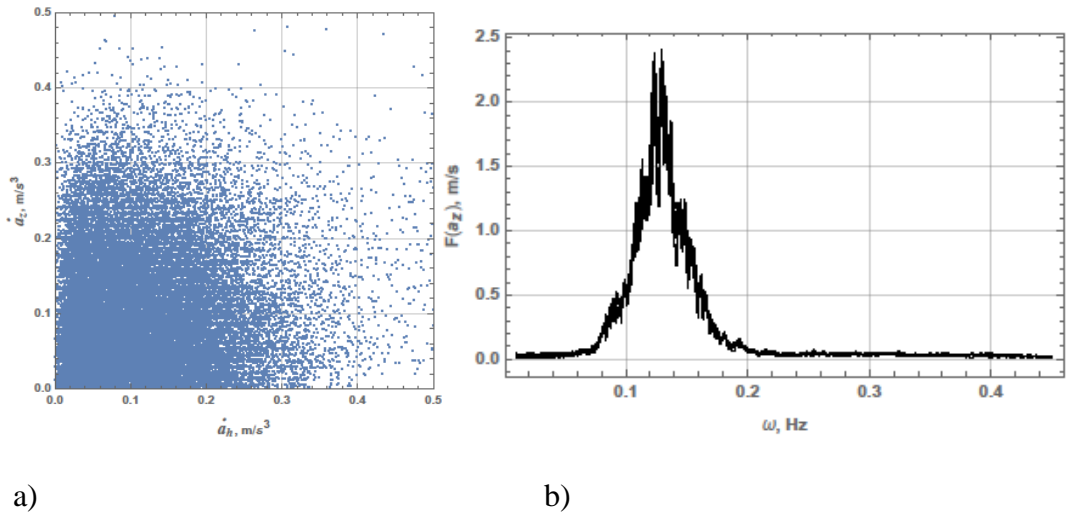

a)  b)

Figure 9. (a) Coordinates of points are equal to the rates of normal and tangential to the floes accelerations, and (b) spectrums of normal to the floe accelerations measured by the IMUs.

## 5. Analysis of SAR image

A study of the evolution of the wave spectrum in the marginal ice zone of the Barents Sea was carried out using a radar (SAR) image taken from the Earth satellite (COSMO-SkyMed) on May 06, 2016. This image was captured in Stripmap mode with a nominal spatial resolution of 3 m and a single polarization (HH). A fragment of the SAR image (see Fig. 10) covers the water area located between the ice edge and the zone of solid (compact) ice with 10/10 concentration at about 78° N and 25.38° E (station 3 in Fig. 1). Waves appear on the picture as alternating bands of light and dark shades.

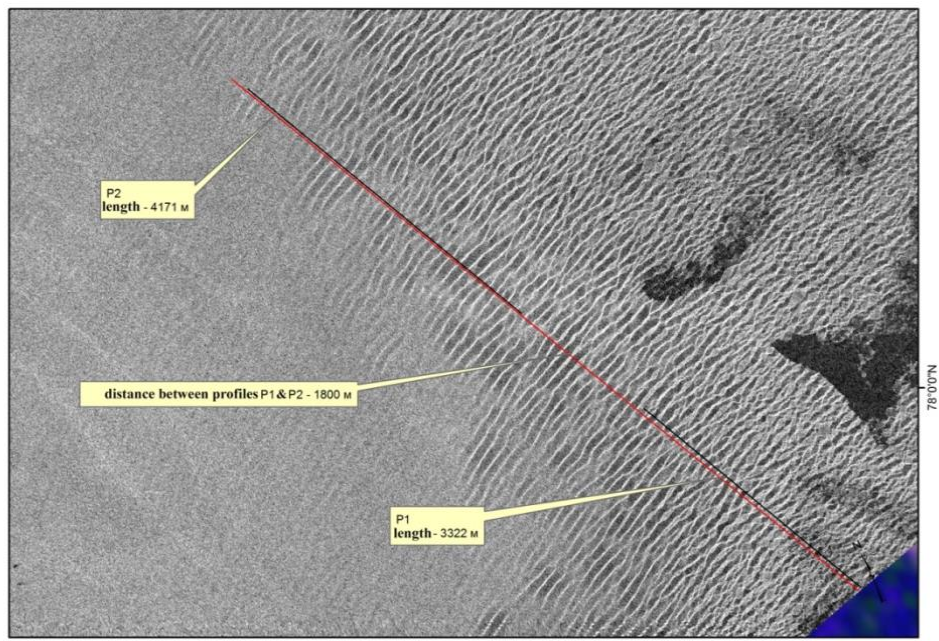

Figure 10. Fragment image from satellite COSMO-SkyMed on May 06, 2016. Black profiles P1 and P2 were used to calculate wavelengths. Red line depicts direction of wave

propagation. The arrow shows the direction of wave propagation.

According to the National Center for Environmental Prediction (NCEP) Climate Forecast System Reanalysis (CFSR) hindcast, on May 06, 2016, around 77 N, 28 East, the wave period was 7.7 s and the significant wave height reached 2.35 m. Assuming that sea depth is of $H = 160$ m, we found the wave number $k = 0.068$ m$^{-1}$ and the wavelength of 92 m. The direction of wave propagation was 155° to the northwest. It is assumed that waves have similar characteristics near the ice edge on the image in Fig. 10.

The wave spectrum is processed by the spectral analysis of the values of the backscattering coefficient that makes up the amplitude part of the SAR image (Marchenko and Chumakov, 2017). The image contains information on the backscattering coefficient ($\sigma^0$) and the phase of electromagnetic pulses scattered back by each object on the scanned surface within the spatial resolution element of the sensors used. The $\sigma^0$-values correspond to the amplitude part of the SAR-snapshot, and the phase values correspond to the phase part. Sigma nought is the conventional measure of the strength of radar signals reflected by a distributed scatterer expressed in deciBels (dB). It is a normalised dimensionless number, comparing the strength observed to that expected from an area of one square meter. Sigma nought is defined with respect to the nominally horizontal plane, and in general has a significant variation with incidence angle, wavelength, and polarisation, as well as with properties of the scattering surface itself (https://sentinel.esa.int/web/sentinel/user-guides/sentinel-1-sar/definitions). The roughness of the sounding surface significantly affects $\sigma^0$. Therefore, the amplitude part is used to obtain information about the scanned surface.

Reflection of radar pulses depends on the surface roughness (Onstott, 2004). Smoother surfaces depict in pictures as dark regions of different intensity and correspond to relatively small values of $\sigma^0$. Surfaces with higher roughness correspond to the areas with large values of $\sigma^0$. Waves on sea surface covered by broken ice appear on SAR-images as alternating lighter and darker bands. The surface roughness is modulated by the wave phase such that their crests correspond to local maxima of $\sigma^0$, and wave troughs correspond to the local minima. The wavelength is determined by the distance between neighbor local maxima of $\sigma^0$ on the wave beam, and equals to a product of the number of pixels between neighbor local maxima of $\sigma^0$ and pixel dimension expressed in meters. Pixel dimension is a constant specified for the SAR image.

Since the process of wave propagation is periodic in nature, we assume that the location of the local maxima $\sigma^0$ on the wave beam is determined by some periodic function, for finding which is permissible to use the Fourier transform. The algorithm for finding the wavelengths using the SAR-image data consists in the following:
1) build using the amplitude part of the image geolocalized raster, containing values $\sigma^0$, dB and to record it in format GeoTif;

2) display in the ArcGis environment the built geolocalized raster in the form of a monochrome image (image);

3) build wave beam in the form of a graphical object in ArcGis;

4) select all the pixels of the image through which the wave beam passes, using the ArcGis tool (3D Analyst) and for each selected pixel determine the value of $\sigma^0$ and the distance from the beginning of the wave beam expressed in the number of pixels;

5) construct a two-dimensional array, according to these data (see section 4), the first column which contains the distance from the origin of the beam expressed in the number of pixels, and the second the corresponding values $\sigma^0$, and export it to Excel;

6) use the average filtered values of $\sigma^{0;}$ to reduce the effect of speckle noise;

7) since all pixels have the same spatial dimensions, the resulting array of values $\sigma^0$ can be considered as temporary series and therefore it is permissible to apply the standard decomposition into the Fourier series for determination the spectral density;

8) determine the periods corresponding to local maxima of the spectral density, and calculate wavelengths expressed in the number of pixels.

The algorithm is applied to various segments (profiles) of wave beam to study changes in the wave spectrum as waves penetrate into the ice covered region. Profiles P1 and P2 shown in Fig. 10 are used to compare the wave spectra near the ice edge and at a distance of 5 km from the ice edge in the direction of wave propagation. The spectral density is calculated as the inverse Fourier Transform from the signal variance function depending on the distance along profiles P1 and P2. The signal variance is measured in $dB^2$, and the distance is measured in the number of pixels. Respectively, the spectral density has dimension of $dB^2 \cdot px$, where px denotes the number of pixels. Figure 11 shows the spectral density versus the wave length expressed in the number of pixels. The dimension of one pixel equals to 1.26 m for the SAR image shown in Fig. 10. The spectral density $\sigma^0$ has spectral maximum on the profile P1 at wave length of 108 m, and the spectral maximum of the profile P2 is located at wave length of 121m. The length of most visible on the SAR-image wave on the distance of 5 km from the ice edge is greater on 13 m than the length of most visible wave on the open water near ice edge.

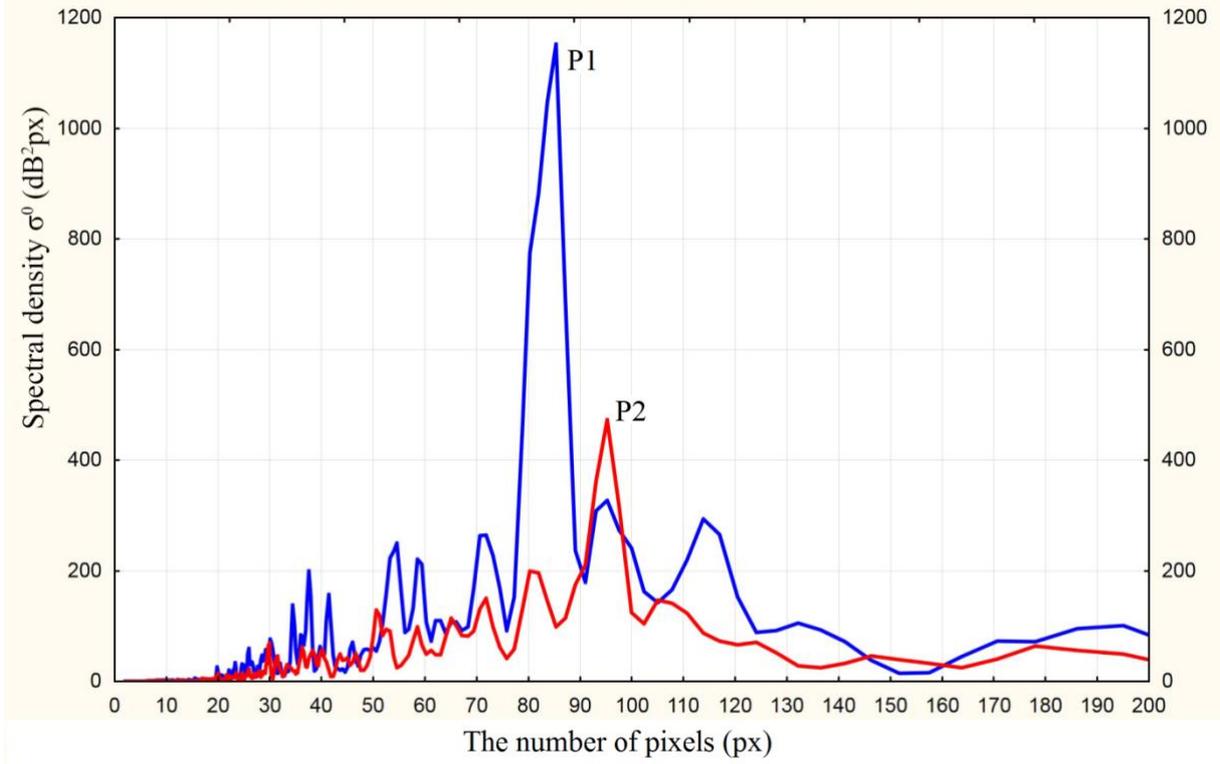

Figure 11. Spectral densities σ⁰ (dB²·px) of the profiles P1 (blue line) and P2 (red line) versus the wave length (px).

### 6. Model of wave damping in broken ice

Momentum balance of water layer with broken ice is written in the form

$$\rho_i h_i \frac{\partial v_i}{\partial t} = -\rho_i h_i g \frac{\partial \eta}{\partial x} - \tau + \frac{\partial \sigma}{\partial x}, \tag{7}$$

where $\rho_i$ and $h_i$ are the ice density and thickness, $v_i$ is the horizontal velocity of ice, $\eta$ is the sea surface elevation, $\tau$ is the drag force between the ice and water, $\sigma$ is the stress in water layer with ice, and $g$ is the gravity acceleration.

The drag force is specified by the formula

$$\tau = \rho_w \nu \frac{\partial v_w}{\partial z}, \quad z = 0, \tag{8}$$

where $\rho_w$ is the water density, $\nu$ is the eddy viscosity in the boundary layer below the ice, $v_w$ is the horizontal velocity of water, and $z$ is the vertical coordinate directed upward. The value $z = 0$ corresponds to the ice bottom.

The horizontal water velocity is a sum of wave induced velocity $v_{ww}$ and boundary layer velocity $v_{bl}$

$$v_w = v_{ww} + v_{bl}. \tag{9}$$

The boundary layer velocity satisfies to the boundary layer equation (Landau and Lifshitz, 1988)

$$\frac{\partial v_{bl}}{\partial t} = \nu \frac{\partial^2 v_{bl}}{\partial z^2}. \tag{10}$$

Solution of equation (7) should satisfy to the non-slip boundary condition below the ice

$$v_i = v_{ww} + v_{bl}, z = 0, \tag{11}$$

and decays outside the boundary layer

$$v_{bl} \to 0, z \to -\infty. \tag{12}$$

The surface elevation and water velocity caused by linear gravity wave in the water with free surface are determined by the formula

$$\eta = a \cos\theta, \quad v_{ww} = v_{ww}^0 \cos\theta \frac{\cosh(k(z+H))}{\cosh(kH)}, \quad v_{ww}^0 = -\frac{\omega a}{\tanh(kH)}, \tag{13}$$

where $a$ is the wave amplitude, $\omega$ and $k$ are the wave frequency and wave number, $\theta = kx + \omega t$ is the wave phase, and $H$ is the water depth. The wave frequency and wave number satisfy to the dispersion equation

$$\omega^2 = gk \tanh(kH). \tag{14}$$

Wave induced trajectories of floating ice and water particles inside the boundary layer are described by equations

$$\frac{dx_i}{dt} = v_i, \frac{dx_w}{dt} = v_{ww} + v_{bl}, \frac{dz}{dt} = \frac{\partial \eta}{\partial t}, \tag{15}$$

where the first two equations describe temporal changes of horizontal displacements of ice and water particles from the time, and the third equation describes temporal changes of their vertical displacements.

In case of periodic sinusoidal wave with small amplitude solution of equations (15) is approximated by ellipses. In case of water with open surface the major axis of the ellipses are horizontal, the minor axis are vertical, and their dimensions decrease with the distance from the water surface (Lamb, 1975). In case when the water surface is covered by ice the ellipses are deformed due to water motion in the boundary layer. It is assumed that the thickness of wave induced oscillating boundary layer $h_{bl}$ is much smaller than the inverse wave number $k^{-1}$. In this case trajectories of water particles located near the boundary layer ($z \approx -h_{bl}$) are similar to the trajectories of surface water particles in the water with open surface (SWT). Trajectories of water particles located near the ice bottom ($z \approx 0$) are similar to the ice trajectories (IT). Therefore, trajectories of water particles inside the boundary layer are located between IT and SWT.

It is assumed that vertical displacement of ice and water are similar within the boundary layer. It means that the vertical axis of IT is equal to $2a$, where $a$ is the wave amplitude. Dimension of the horizontal axis of IT depends on the interaction between floes. If the floes are in close contact (compacted ice) then their horizontal displacements are confined and the horizontal axis of IT ellipse is smaller than the vertical axis. In case of diffuse ice floes don't interact with each other and the horizontal axes of IT and SWT are similar. Solid black and blue lines show IT and SWT trajectories in Fig. 12 for compacted ice (a), less compacted ice (b) and diffuse ice (c). Dashed blue lines correspond to the trajectories of water particles inside the boundary layer.

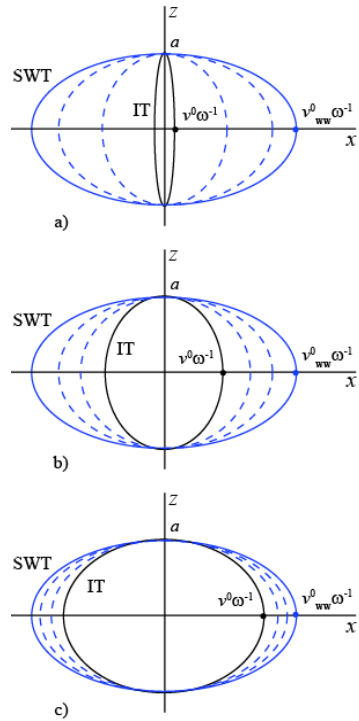

Figure 12. Trajectories of ice (inner solid black lines) and water particles (outer solid blue lines) induced by periodic waves with small amplitude within the oscillating boundary layer: (a) compacted ice, (b) less compacted ice, and (c) diffuse ice.

Wave energy dissipates in the boundary layer when waves propagate below the ice. It is assumed that the energy dissipation over the wave period is much smaller the wave energy. In this case the dispersion equation accounting wave damping effect has the form

$$\omega_d = \omega(1 + i\beta),\ 0 < \beta \ll 1, \qquad (16)$$

where $\beta$ is a damping coefficient.

To calculate the change of the wave number $\delta k$ when the wave with wave number $k$ penetrates from open water region below the ice, it is assumed that the wave frequency $\omega$ is conserved. From (16) it follows

$$\omega'\delta k + \omega''\delta k^2/2 + i\omega\beta + \omega\beta^2 = 0, \tag{17}$$

where $\omega'$ and $\omega''$ are the first and the second derivatives from $\omega$ with respect to $k$. The solution of (17) is approximated with the formulas

$$\delta k = i\beta k_1 + \beta^2 k_2 + O(\beta^3), \; k_1 = -\omega/\omega', \; k_2 = k_1 + \omega^2\omega''/2\omega'^3, \tag{18}$$

where $(\beta k_1)^{-1}$ is the reference e–folding distance of the wave amplitude, and $\beta^2 k_2$ is the reference wave number change related to solid flexible ice. The wave length increases due to the damping effect since $\omega'' < 0$, but this effect is very small because $\beta \ll 1$.

The wave energy dissipation rate over the unit area is calculated by the formula

$$D' = -\rho_w \nu \int_{-\infty}^{0} \left(\frac{\partial v_{bl}}{\partial z}\right)^2 dz. \tag{19}$$

The attenuation rate of the wave amplitude due to the friction on the water surface is

$$\frac{\langle D' \rangle}{2E} = -\omega\beta, \; 2E = \rho_w g a^2, \tag{20}$$

where $E$ is the wave energy per unit area of water surface, and $\langle D' \rangle$ is the wave energy dissipation rate averaged over the wave period. The decay of the wave amplitude is described by the factor $\exp(-\omega\beta t)$.

Periodic solution of equation (10) with boundary condition (12) is written in the complex form

$$v_{bl} = v_{bl}^0 e^{i\theta + \lambda z}, \; \lambda = (1+i)/h_{bl}, \; h_{bl} = \sqrt{2\nu/\omega}, \tag{21}$$

where $h_{bl}$ is the thickness of oscillating boundary layer below the ice, and $v_{bl}^0$ is a complex amplitude of the boundary layer velocity. Assuming that the ice velocity is written in the complex form as $v_i = v^0 e^{i\theta}$, where $v^0$ is a complex amplitude of the ice velocity, we find from boundary condition (11)

$$v_{bl}^0 + v_{ww}^0 = v^0. \tag{22}$$

Formula (22) is satisfied when

$$\frac{v_{bl}^0}{v_{ww}^0} = -\alpha, \; \frac{v^0}{v_{ww}^0} = 1 - \alpha, \tag{23}$$

where $\alpha$ is a complex coefficient. The absolute value of $1-\alpha$ has clear physical meaning: it is a ratio of the ice velocity amplitude to the water velocity amplitude caused by the periodic wave. In compacted ice $\alpha \approx 1$ and in diffuse ice the absolute value of $\alpha$ is close to 0 (Fig. 12).

From equation (7), it follows that the coefficient $\alpha$ is related to the complex amplitude of ice stresses $\Omega$ by the formula

$$\Omega = \alpha g a (\rho_i h_i + \rho_w (1+i) h_{bl}/2). \tag{24}$$

Substituting the real part of the complex velocity $v_{bl}$ determined by formula (21) into formulas (19) and (20) we find

$$\langle D' \rangle = -\rho_w g \frac{\sqrt{\nu\omega}}{2\sqrt{2}} \frac{k|\alpha|^2 a^2}{\tanh(kH)}, \quad \beta = \frac{h_{bl} k |\alpha|^2}{4 \tanh(kH)}. \tag{25}$$

The thickness of oscillating boundary layer and the wave damping coefficient are shown in Fig. 13 versus the wave frequency. It is calculated with three different values of the eddy viscosity covering the range of the eddy viscosities shown in Fig. 8b. The thickness of oscillating boundary layer increases with increasing of the eddy viscosity and decreases with increasing of the wave frequency. The wave damping increases with increasing of wave frequency and eddy viscosity. This effect influences significant changes in the wave spectrum as waves propagate into the ice covered region. Further, it is shown that the frequency of the most energetic waves shifts to lower frequencies due to preferential damping of high frequencies.

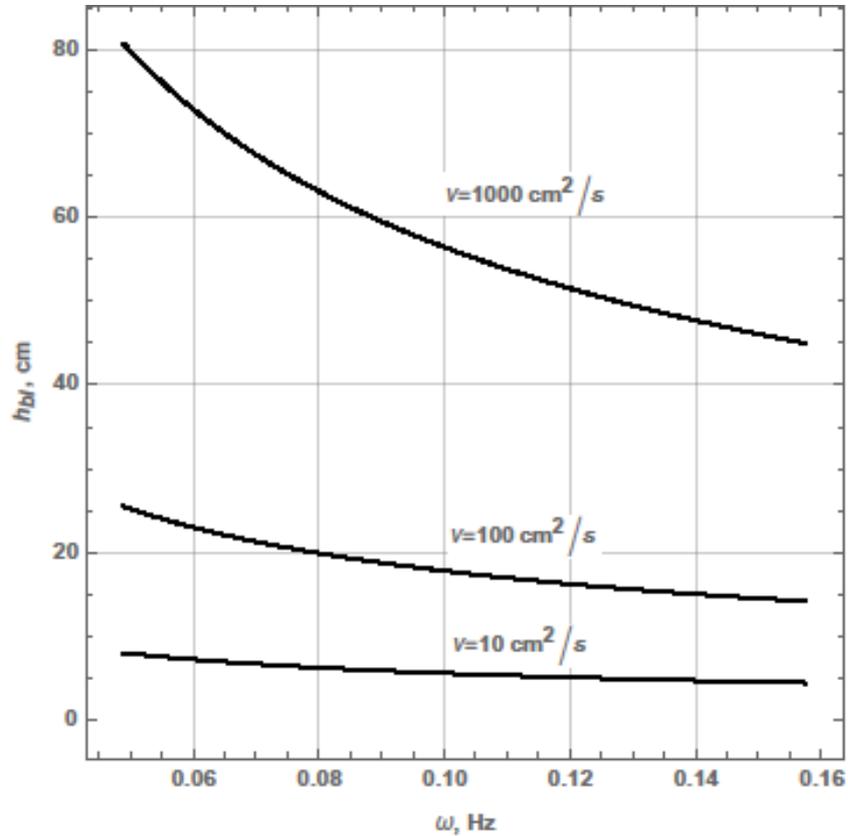

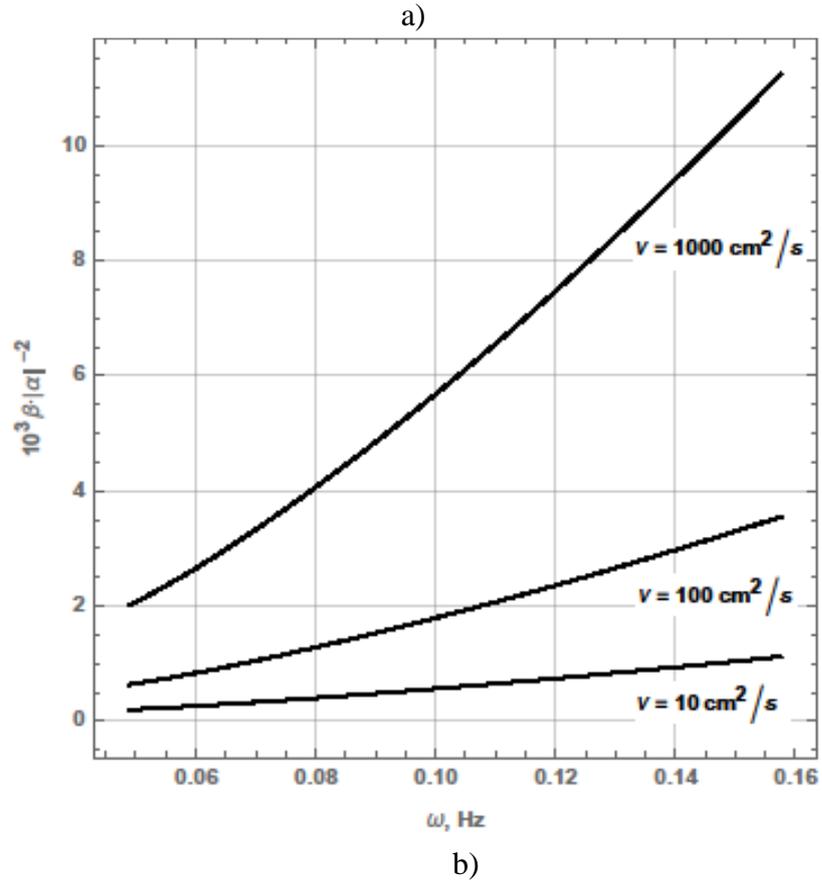

a)

b)

Figure 13. The thickness of the oscillating boundary layer (a) and the wave damping coefficient (b) versus the wave frequency calculated with different values of the eddy viscosity.

Wave propagation in MIZ is described by the equation (Perrie and Hu, 1996)

$$\frac{\partial E}{\partial t} - c_g \frac{\partial E}{\partial x} = -2\omega\beta E, \qquad (26)$$

where $c_g = \omega'$ is the wave group velocity. Solutions of equation (26) describing temporal and spatial attenuations of spectra of wave amplitude and energy are constructed respectively with the assumptions $\partial/\partial x = 0$ and $\partial/\partial t = 0$. They are described by formulas

$$\eta_f = \eta_{f0}(\omega)e^{-\omega\beta t}, \; E_f = E_{f0}(\omega)e^{-2\omega\beta t}; \; \eta_f = \eta_{f0}(\omega)e^{-\omega\beta c_g^{-1} x}, \; E_f = E_{f0}(\omega)e^{-2\omega\beta c_g^{-1} x}, \qquad (27)$$

where $\eta_{f0}$ and $E_{f0}(\omega)$ are the initial values of the spectra of wave amplitude and wave energy by either $t = 0$ (temporal attenuation) either by $x = 0$ (spatial attenuation).

The wave damping coefficient depends on two parameters characterizing broken ice: the eddy viscosity in the boundary layer below the ice and the coefficient $\alpha$. The coefficient $\alpha$ depends on the internal stresses in the ice and proportional to the ratio of the ice velocity amplitude to the amplitude of wave induced water velocity. If the ice velocity is similar to the wave induced water velocity then $\alpha = 0$ and wave energy dissipation is absent. If the ice velocity equals to zero then the wave energy dissipation rate determined by formula (25) coincides with the formula derived by Weber (1987). The coefficient $\alpha$ can be estimated using the records of accelerometers deployed on the ice.

As an example of the model application we describe the effect of wave spectra shift described in previous section. It is assumed that spectrum $E_{f0}(\omega)$ is given in the form of the JONSWAP spectrum on open water near the ice edge (see, e.g., WMO, 1998), the wind speed is 10 m/s, the wave fetch is 200 km and the sea depth is $H = 160$ m. The eddy viscosity equals $\nu = 100$ cm²/s. Figure 14a shows the evolution of the wave spectra $E_f(\omega)$ with the increase of the distance from the ice edge, where the frequency $\omega$ is related to the wave number by dispersion equation (14). According to the estimates in Section 3 ice effect of the wave dispersion is small and can be ignored. Figure 14b shows wave lengths $\lambda_{max}$ corresponding to the wave numbers of most energetic waves versus the distance from the ice edge. The increase of the length of most energetic wave exceeds 10 m over 5 km distance when $|\alpha| = 1$. This value is close to the shift of wave length calculated by the analysis of spectral density ratio $\sigma^0$ (Fig. 11).

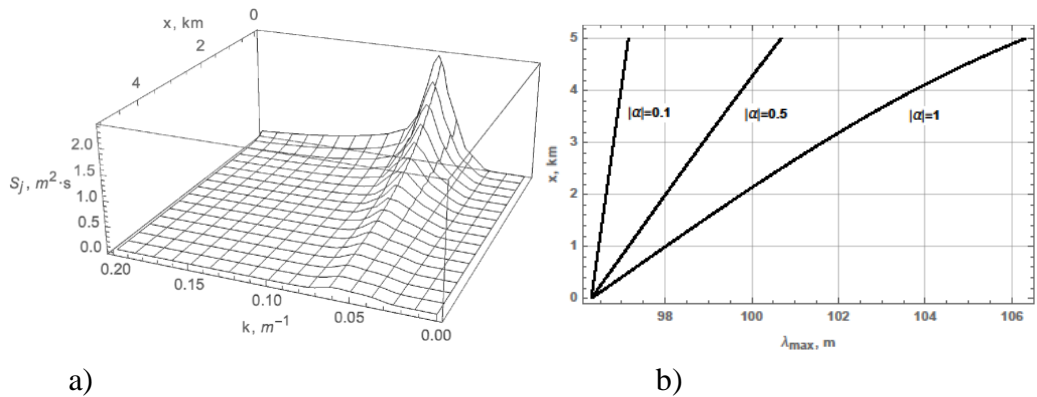

a)          b)

Figure 14. Spectral density of wave energy versus the distances from the ice edge and wave number calculated with $|\alpha| = 1$ (a). The length of most energetic wave versus the distance from the ice edge (b).

## 7. Discussion and conclusions

The marginal ice zone in the North-West Barents Sea is not stable and its configuration changes depending on the action of wind drag. Wind waves and swell penetrate into ice covered region from Atlantic Ocean and from open water areas of the Barents Sea. This property influences different wave characteristics in different parts of the MIZ over relatively small distances depending on their relative geographical locations with respect to Spitsbergen and wave fronts in the open sea. In the present paper this effect was discovered and analyzed.

Wave observations on Station 1, located at 25.5°E, 77.76°N, showed transformation of wave spectrum around 23:00 – 24:00 of May 01, 2016. The frequency of most energetic waves shifted in this time from the frequency of 0.125 Hz (8s period) to the frequency of 0.09 Hz (11 s period). At the same time drift ice was displaced by the West wind to the East from Edgeøya (Fig. 2b). The change in location aligned the floe with an area of low ice concentration allowing penetration of swell from South-West to North-East along Edgeøya (Fig. 1, May 02).

Characteristics of the wave spectrum measured on Station 1 were compared with periods and directions of the propagation of most energetic waves calculated with the Climate Forecast System Reanalysis (CFSR) hindcast. The simulation results are available on the web (https://earth.nullschool.net). It is assumed that waves registered in Station 1 propagated from the South directions. The data from 6 points located at 77ºN latitude with longitudinal coordinates of 22ºE – 27ºE were used. The points are shown in Fig. 1 (May 02). Periods and directions and of most energetic waves are shown respectively in Fig.15. Figure 16 shows the dependence of significant wave heights in these points versus the time together with significant wave heights shown in Fig. 6b.

Figure 15a shows that the period of most energetic waves increased from 4 s to 10 s after 00:00, May 02, in the points with coordinates of 22ºE and 23ºE. The direction of wave propagation in the point with coordinate of 22ºE was 20º to the North-East on 02:00 and 04:00, May 02 (Fig. 15b). The direction of wave propagation in the point with coordinate of 23ºE changed from 165º from the South-East on 02:00 to 35 º to the North-East on 04:00, May 02 (Fig. 14b). Periods of waves in the points with coordinates of 24ºE – 27ºE were slightly above 6 s on the May 01 and May 02 (Fig. 15a). Direction of wave propagation in these points was determined to be around 170º, from a South-South-East direction (Fig. 15b). The direction of wave propagation in the point with coordinate of 22ºE on May 02 corresponds to the direction of wave propagation shown in Fig. 7b.

The period of most energetic waves (8 s) measured on station 1 is greater than two times the period of simulated waves in the points with coordinates of 22ºE and 23ºE on May 01 (Fig. 15b). Measured period of most energetic waves is 11 s, and period of simulated waves is 10 s on May 02. Thus, the shift of wave periods between simulated and measured waves was greater for smaller periods. This may correspond to the property of stronger damping of waves with shorter periods.

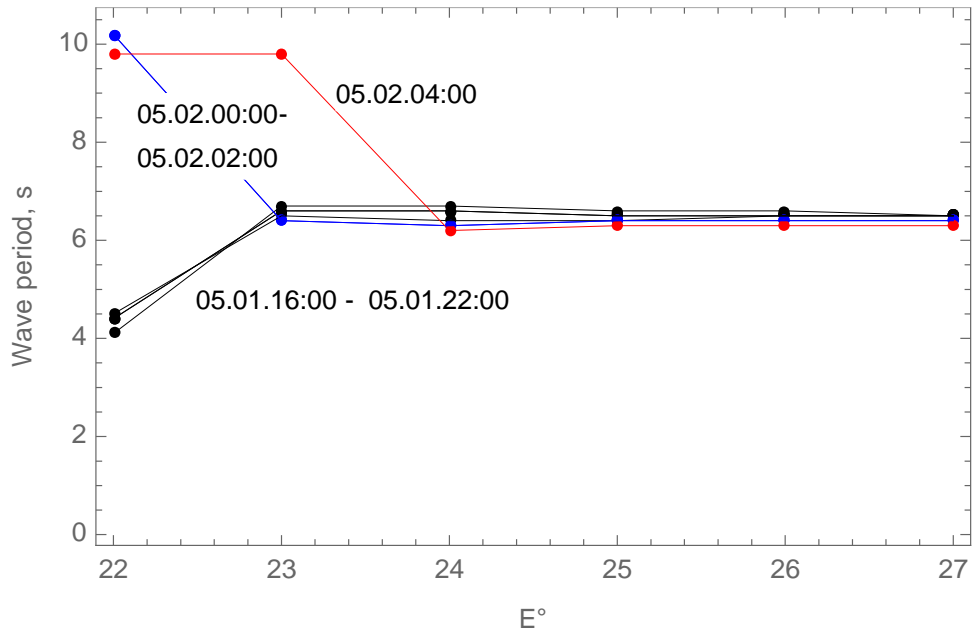

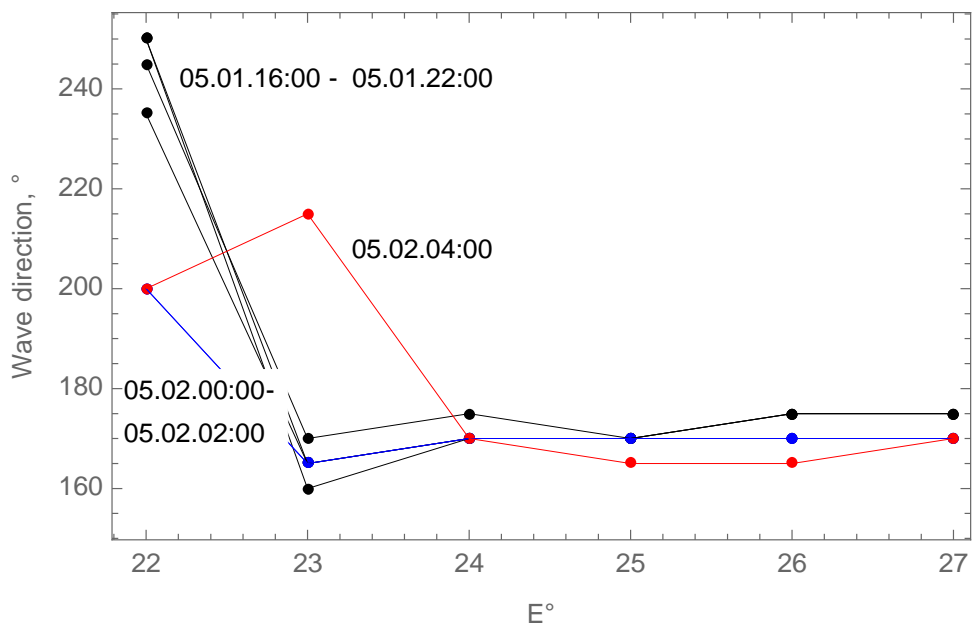

Figure 15. Periods (a) and directions (b) of most energetic waves at 77°N versus the longitude from the CFSR hindcast from 16:00 to 24:00, May 01, are shown by black lines. Blue and red lines show the same quantities respectively on 02:00 and 04:00 of May 02, 2016.

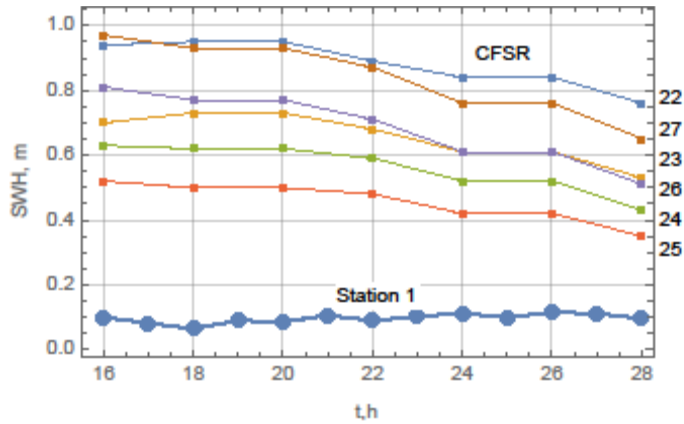

Figure 16. Thin lines show significant wave heights of most energetic waves at 77°N versus the time from the CFSR hindcast from 16:00, May 01, to 04:00, May 02. Thick blue line shows significant wave heights calculated from the field data recorded on Station 1. Numbers from the right side of the frame correspond to the longitudinal coordinates of 6 points located at 77°N latitude and shown in Fig.1 (May 02).

Figure 16 shows visible reduction of significant wave heights on the way of their propagation from open water at 77°N to Station 1 on May 1 and May 2. Temporal changes of significant wave heights provided by the CFSR hindcast are stronger than temporal changes of significant wave heights calculated from the field data on Station 1. Time gradients of these data have different signs on May 01, and the same sign from 02:00 to 04:00, May 02. It can be explained by strong damping of most energetic waves with periods of 4 s and 6 s provided by the hindcast data on May 01. As a result, these waves were not registered on Station 1. At the same time, waves with periods of 8 s and greater recorded on Station 1 were not included in the hindcast data on May 01 since their energy on the open water was smaller than the energy of waves with periods of 4 s and 6 s. Correlation between significant wave heights on the open water and Station 1 became better on 02:00 and 04:00, May 02, since periods of most energetic waves (10 s and 11 s) became greater and their damping was smaller.

The effect of wave length increase on the way of wave penetration below the ice was described using the SAR-image with a nominal spatial resolution of 3 m and a single polarization. The image made on May 06 provided information on ice conditions on Station 3 (Fig. 1, May 06) which is distant on few kilometers from Station 1 in the North-East direction. The ice was compressed by the East-South-East wind at this time, and damping of gravity waves propagating from South-East was relatively strong. Increase of the length of most energetic waves reached of about 10 m over a 5 km distance.

Measurements performed on broken ice on Station 2 (Fig. 1, May 04 and May 06) showed that period of most energetic wave was of 8 s. The same period was found in the CFSR hindcast for the points at 77°N closed to Station 2. Very small damping of surface waves in the region was also discovered by direct measurements with IMUs distributed over 5 km distance extended along the direction of wave propagation (Tsarau et al., 2017). The small damping was also confirmed by the distribution of the acceleration rates of floes shown in Fig. 9a on the plane of horizontal and vertical acceleration rates. They fill range with circular envelop corresponding to the acceleration rates of surface water particles accelerated by the waves. Vertical accelerations of the floes are identical to the vertical accelerations of the water particles. Amplitudes of the horizontal accelerations of the water particles are similar to the amplitudes of their vertical accelerations on deep water. Therefore, amplitudes of horizontal accelerations of the floes are similar to the horizontal accelerations of the water particles. Thus, the floes motion was similar to water motion, the relative motion was not significant and energy dissipation was very small. Spectra recorded by IMUs deployed on neighbor floes were very similar (Fig. 9b), but not exactly the same. Sampling frequency of 10 Hz was probably not enough to resolve vibrations from floes collisions. Visually, floe drift was very smooth and it was impossible to distinguish floe motion demonstrating their collisions.

A model of wave damping in broken ice was formulated taking into account wave induced oscillating boundary layer below the ice. According to the model, broken ice has smooth periodical motion similar to the motion of surface particles of the water in waves. In case of compressed ice wave induced horizontal motion of the ice is absent, and in case of loose and not compressed ice the horizontal motion of the ice is similar to the horizontal motion of the water below the ice. Vertical motion of broken ice and water is always similar. Difference between the horizontal motion of the water and ice is characterized by the parameter $\alpha$ in the model. The motion of water and ice is similar when $\alpha = 0$. The wave induced horizontal motion of ice is absent when $\alpha = 1$. The wave damping occurs due to the interaction of waves with under ice turbulent boundary layer created by ice drift. The turbulence was characterized by the eddy viscosity in the model. Formula (25) describing wave energy dissipation coincides with formula (4.15) of Weber (1987) and formula (A11) of Liu and Mollo-Christinsen (1988) derived for the water with infinite depth when $\tanh(kH) = 1$ and $\alpha = 1$. It was mentioned in these papers that the eddy viscosity should be used to describe wave energy dissipation in the oscillation boundary layer, but the eddy viscosity has not measured previously.

The eddy viscosity can be calculated with the standard formula (4) when the shear stresses in the water below the ice are known. The shear stresses can be calculated using formula of Reynolds stresses (see e.g., Landau and Lifshitz, 1988) and high frequency measurements of water velocity fluctuations below the ice. Measurements of under ice velocity with ADV performed on Station 1 showed that the eddy viscosity below drift ice can exceed 100 cm$^2$/s. Correlation between the floe acceleration and the eddy viscosity was discovered. The values of the eddy viscosity are similar to the values of the eddy viscosity calculated by the same method using the field measurements performed in the same place in the other years (Marchenko et al, 2015; Marchenko and Cole, 2017).

Viscous properties of broken ice can also be explained by the influence of slush between and below floes on their motion. The slush is produced due to floes interaction. Viscosity of slush estimated below 300 cm$^2$/s (see, e.g., Newyear and Martin, 1999; Carolis et al, 2005; Rabault et al, 2017) can be higher than the eddy viscosity. The model described in the paper can be used for the description of wave damping in broken ice with slush using higher values of the viscosity $\nu$.

**Acknowledgements**

The author wishes to acknowledge the support of the Research Council of Norway through the SFI SAMCoT, IntPart project AOCEC, and PETROMAKS2 project Dynamics of floating ice.